*Original Article*

# A Framework for Successful Corporate Cloud Transformation

Naga Mallika Gunturu

*IEEE member, IEEE Computer Society Member, Senior Project Manager, Program Manager, Virginia USA*

***Abstract -*** *Corporate Cloud transformation is expected to continue to grow double-digit each of the next few years. This growth is augmented by digital transformation, which itself is gaining huge momentum due to the recent consumer behaviour trends and especially the COVID pandemic. It is also estimated that globally billions of dollars are wasted due to efficiencies in the way cloud migrations are launched and handled. This paper discusses a framework using which organizations can successfully execute cloud transformation.*

## I. INTRODUCTION

Digital transformation is a megatrend that enterprises are experiencing, and the Covid pandemic has only accelerated this. Broadly speaking, digital transformation can be defined as the integration of digital technology into all areas of a business resulting in fundamental changes to how businesses operate and deliver value to customers. Beyond that, it's a cultural change that requires organizations to continually challenge the status quo, experiment often, and get comfortable with failure [7].

At the beginning of the COVID pandemic, digital acceleration was viewed largely as a way to enable remote work and ensure business continuity. Two years later, research indicates that organizations now see digital acceleration as a more permanent fixture in the business landscape. Eighty-six per cent of 326 business executives across different functions surveyed by Harvard Business Review Analytic Services in April 2021 said their organization had accelerated its digital transformation during the pandemic. Of those respondents that rapidly transformed, 91% said they plan to maintain the faster speed—or possibly increase it—even after the pandemic ends [8].

Cloud transformation is an important enabler for enterprises to achieve their digital transformation objectives. The question is no longer whether the cloud should have a place in a company's strategy – the question is about the form and scope of that place [9]. Business agility is the key for enterprises to stay relevant and competitive. Technical agility is necessary to achieve this. Technical agility is the ability to adapt quickly and smoothly or integrate current technologies with newer, different, disruptive, expansive or convergent technologies [11]. Cloud provides organizations to quickly set up and ramp up the infrastructure needed and provides great flexibility to support the organization's operational needs. With the cloud, organizations are able to benefit in many ways - from reduced upfront CAPEX investments to be able to provide products and services to customers anywhere, anytime while having the ability to scale up or scale down based on the demand of the day.

McKinsey projects that by 2024, most enterprises aspire to have $8 out of every $10 for IT hosting go toward the cloud, including private cloud, infrastructure as a service (IaaS), platform as a service (PaaS), and software as a service (SaaS) [6].

Gartner projects that global cloud adoption spending will grow 21.7% to reach $482 billion in 2022. Additionally, by 2026, Gartner predicts public cloud spending will exceed 45% of all enterprise IT spending, up from less than 17% in 2021[10]

Although companies are transitioning more workloads to the public cloud, missteps in coordinating the migration are taking a toll. A study by McKinsey shows that those inefficiencies are costing the average company 14 per cent more in-migration spending than planned each year, and 38 per cent of companies have seen their migrations delayed by more than one quarter [6]. This translates to billions of dollars of unanticipated expenditure and wastage due to delays.

Before going into the details of a framework following which enterprises could ensure successful cloud migration, let us first look into some of the critical success factors without which it is imperative that the cloud transformation fails.

- Executive buy-in and engagement - As mentioned earlier, cloud transformation can positively impact the technical agility and, in turn, the business agility of enterprises. It is but natural that initiatives of this nature are resource and cost-intensive and would need several teams in the enterprise to contribute at various levels. Having an executive alignment and engagement hence becomes a critical success factor for ensuring the necessary buy-in from teams across the organization.
- A well-defined cloud strategy is a critical success factor for cloud transformation as it provides a clear direction of why the organization is going for the cloud transformation and what are the acceptable limits within which the transformation should happen. This information can be used as a guiding north star by teams for any decisions they need to take in due course of the transformation.
- Adherence to security and compliance requirements - Every business must adhere to certain laws and regulations as part of its operations. This is called regulatory compliance. These compliance requirements are determined by state, federal, international, industry-specific regulations. Failure to meet security and compliance requirements will negatively impact the trust that customers place in a business. Cloud transformation introduces new security and compliance risks as the data and operations move out of the exclusively owned enterprise data centres to a shared model. Addressing these risks in the early phases of cloud transformation is crucial to avoid hefty fines and loss of face later.
- Controlling costs - A major concern for enterprises moving to the cloud is unanticipated costs. Though cloud migration benefits the enterprises by reducing the need for hefty upfront CAPEX investments, the pay-per-use model of cloud can quickly spiral out if the necessary controls and guardrails are not put in place.
- Change Management - Cloud transformation brings with it many changes: technology, business, people, processes. An effective change management process will avoid chaos, panic and set a precedence for acceptance of the change.

## II. CLOUD MIGRATION FRAMEWORK

Enterprise cloud migrations are typically multi-year initiatives. There is a lot of material available on how organizations should decide on the cloud model, security considerations etc. However, there is not a lot of information available detailing a cohesive framework that an enterprise could follow to launch and successfully execute cloud migration. This paper aims to cover this gap and proposes the following 5-phase framework.

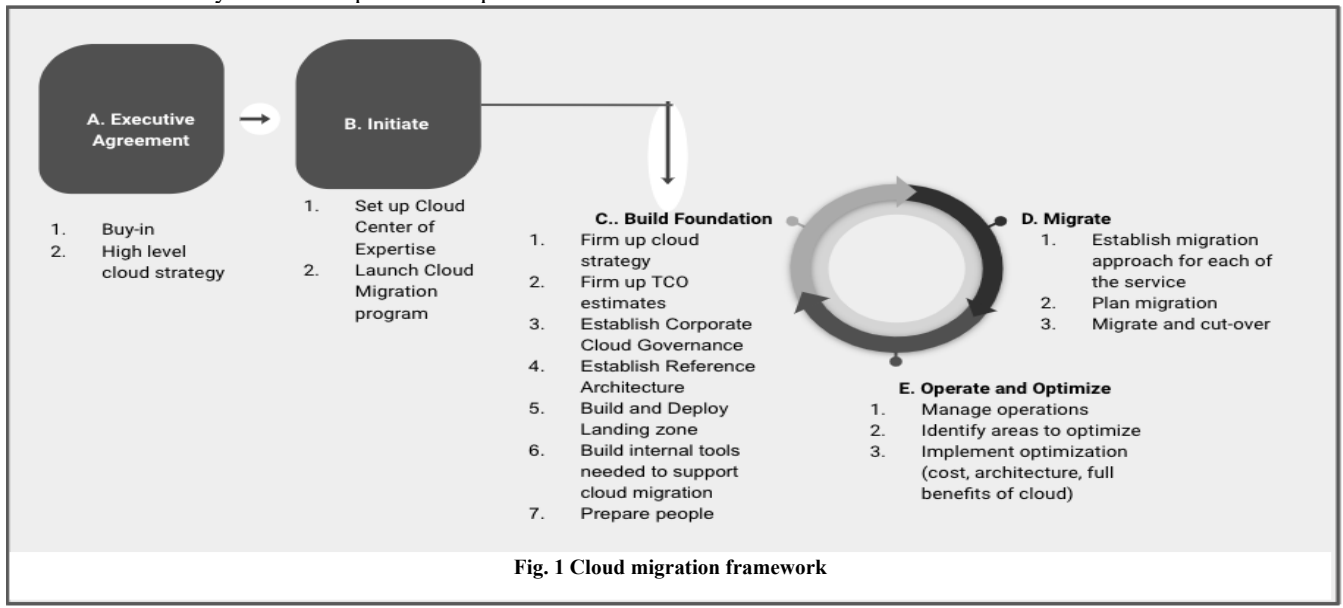

**Fig. 1 Cloud migration framework**

### A. Executive Direction

Every enterprise cloud migration initiative should start with a solid high-level assessment. The outcome of this phase is that executive leadership decides on whether the organization should embark on a cloud migration journey. And if it decides so, executive sponsorship is established. Key questions that should be considered in the evaluation of this decision are:

- Why should we move to the cloud?
- What business and technical outcomes does the organization want to achieve by migrating to the cloud?
- Which applications and/or infrastructure are being considered to move to the cloud?
- What is the organization's readiness in terms of people and processes?
- How does cloud transformation align with the organization's corporate and business strategies?
- What other strategic priorities will cloud migration compete with?
- What is the high-level timeline within which cloud migration is expected to happen?
- What is the rough budget estimate needed to migrate and operate services in the cloud?

Establishing a clear understanding of the above questions helps to establish a strong business case, take a rational decision and rally the entire organization behind the decision to migrate to the cloud. It is important to note that an enterprise embarking on a cloud journey should prepare not only for adapting itself to technology changes but also prepare its people to embrace a mindset suitable for building and operating services in the cloud.

To ensure sustained executive engagement, a cloud strategy team consisting of leaders from business lines, IT, Operations, Finance, Legal, Security, Architecture, Infrastructure, which validates and maintains alignment between business priorities and cloud adoption efforts throughout the migration journey, is established.

*B. Initiate*
Once executive alignment and sponsorship are established, the next step is to set in motion a set of actions that lead to a strong foundation. It starts with establishing a Cloud Center of Expertise (CCOE), which is a centralized team led by the enterprise architecture function that leads and governs cloud computing adoption within an organization [2]. CCOE's responsibilities are:

- Establish Corporate Cloud Governance
- Set direction in selecting cloud providers and architect the cloud solution(s)
- Collaborate with the procurement team for contract negotiation and vendor management
- Be Cloud Knowledge Owner and raise the level of cloud knowledge of others within the organization

Further, at this stage, a Cloud migration program with the following workstreams should be launched. The objective of this program is to drive the migration of workloads and applications to the cloud along with the necessary environments, processes and tools to monitor and operate the services in the cloud.

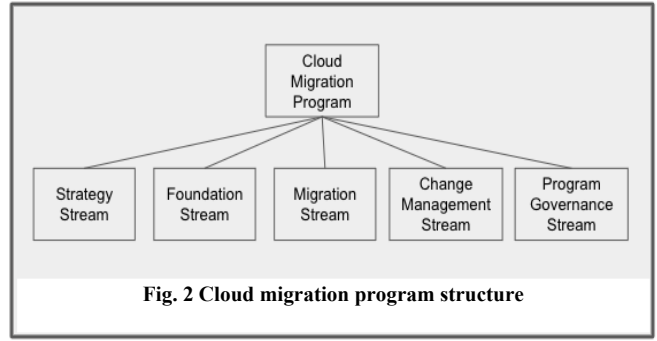

**Fig. 2 Cloud migration program structure**

- The strategy stream focuses on defining and refining the strategy. As you will see, while a high-level corporate cloud strategy is defined during the executive agreement phase, it gets refined and more concrete in further phases
- The Foundation stream focuses on putting the foundation blocks in place so that the cloud migration can be accelerated with the necessary guidelines and guardrails in place
- The migration stream focuses on the migration implementation activities
- The change Management stream focuses on people and processes that should be prepared and adopted for the changes that cloud migration would bring in
- The program Governance stream focuses on delivering the program objectives on time and within budget. This workstream spans all phases of the program to first establish the program organization and later execute, assess, ensure and report the progress of the cloud transformation journey.

Program launch must start with a kick-off meeting conducted by the program manager involving the cloud strategy team, CCOE, development, and operations lead, along with participants from ancillary support teams like procurement, finance and legal.

*C. Build Foundation*
This phase focuses on getting the foundation blocks in place through the strategy, foundation, change management and program governance streams of the program.

*a) Program Governance workstream*
At this stage of the program, this workstream emphasizes setting in place a number of things that define how the program will be managed.

*1) A communication plan*
A communication plan is established**.** This entails details of engagement with the contributing teams and the escalation path in case of impediments. This plan also covers how communication is propagated to the stakeholders. Further, this plan also details the cadence for program meetings.

*2) Change Management Plan*

Change Management plan covers procedural, training, communication aspects to enable the organization's readiness to accept and embrace the changes that will be brought in the wake of cloud transformation.

*3) Schedule Management*

Schedule management primarily focuses on the immediate deliverables expected within a quarter and a long-term schedule view of the overall migration. It is important to note that the planning should be done in a rolling wave fashion to accommodate for the learnings that teams will make as they progress on the cloud journey.

*4) RACI matrix*

RACI matrix covering roles and responsibilities of participating teams

*5) Reporting framework*

Reporting framework comprising program dashboard, financial reporting, risk and issue register and decision log.

Program Governance workstream will start small and will expand and fine-tune as progress is made and strategy and other elements become clearer. For instance, at the start, vendor management is not part of the program governance, but once a decision is taken on the cloud service provider and cloud service integrator, a plan for managing the related activities and deliveries will be added to the program management plan.

*b) Strategy workstream*

The CCOE team firms up the cloud strategy by addressing important questions like "what cloud model will the organization adopt - single, multiple or hybrid?", "Which cloud service provider(s) will the organization use?", "Will the organization engage a cloud migration service integrator?", "What is the disaster recovery/business continuity model that is applicable for the services that migrate to the cloud?". An inventory of the current processes and services along with inter-dependencies must be established to be able for establishing a firm strategy. At this early stage, it could also be beneficial to reach out to peers in the industry and learn best practices from their experiences. An analogy to note here is that of agile transformation. While there are standard agile principles and methodologies defined, each industry and organization implements a nuanced approach that gets refined over a period of time. Generally, businesses within the same industry have agile implementations aligned to a large extent and can greatly learn from the success or failure of each other. Similarly, businesses within the same industry can learn from each other and strengthen their cloud transformation success rate.

TCO (Total cost of ownership) should be firmed up once the key aspects that impact the cost of the migration, such as the choice of the Cloud Service Provider, choice of the Cloud Service Integrator, licensing model and cloud services needed, are established. This information should be taken back to the cloud strategy team to validate whether the cost justifies the business case.

*c) Foundation workstream*
*1) Establish Corporate Cloud Governance*
As part of this workstream, CCOE creates corporate cloud governance policies and selects governance tools in collaboration with a cross-functional team [1]. Cloud governance should strike a balance between business drives and requirements, risks, and the existing compliance requirements. The existing policies and security practices must be extended for the cloud environment.

The Open Group defines the following cloud governance principles that should be applied throughout an organization's cloud journey [13]:

- **Principle 1:** Compliance with Policies and Standards - Cloud standards should be open, consistent with, and complementary to standards prevalent in the industry and adopted by the enterprise.

- **Principle 2:** Business Objectives Must Drive Cloud Strategy - Enterprise cloud strategy should be an integral part of the overall business and IT strategy driven by both the "business of the business" and the "business of IT" objectives for the enterprise.

- **Principle 3:** Collaborative Contracts Between Citizens of the Cloud Ecosystem - A clear set of rules and agreements that define the interaction between stakeholders is essential for enabling their healthy coexistence within the cloud ecosystem.

- **Principle 4:** Adherence to Change Management Processes - Change should be exercised and enforced in a consistent and standardized manner across all the constituents in the cloud ecosystem of an enterprise.

- **Principle 5:** Enforcement of Vitality Processes to Achieve Continuous Improvement - Cloud computing governance processes must dynamically monitor events that trigger continuous improvements.

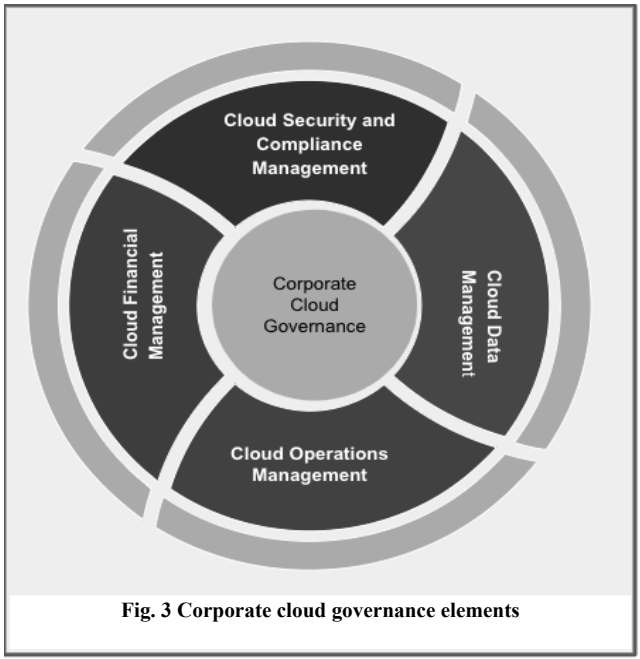

**Fig. 3 Corporate cloud governance elements**

A Corporate Cloud Governance should cover the following at a minimum:
- Cloud Security and Compliance Management control to enable enterprises to leverage the benefits of moving services and infrastructure to the cloud while minimizing the risks of using third party infrastructure and software. These controls cover the areas of user authentication and authorization, data encryption, data loss prevention, vulnerability management, regulatory requirements.
- Cloud Data Management controls encompass policies and procedures that give enterprises control of their business data, both in the cloud and in setups where data is stored or sourced in a combination of on-premises and cloud applications [4]. These controls cover the areas of data classification, data access controls, encryption policies for data at rest and data in transit, data life cycle management like archival, purge, backup.
- Cloud Operations Management controls cover the policies and processes for enterprises to successfully operate in the cloud. These controls cover areas of configuration and software deployment, provisioning and orchestration, access controls to the operators, the definition of SLAs and monitoring of SLA compliance, health check monitoring, performance and capacity monitoring, logging, synthetic monitoring,
- Cloud Financial Management controls cover the policies and processes to be put in place, so that budget overruns are prevented, and optimal utilization of finance spent on cloud usage is achieved. This covers the financial policies that set the direction for cloud adoption of services, budget limits and consistent cost reporting.

*2) Establish Reference Architecture*

OpenGroup defines reference architecture as a generic architecture that provides guidelines and options for making decisions in the development of more specific architectures and the implementation of solutions [5]. CCOE establishes Cloud Reference Architecture which is a framework that all cloud migrations in the organization must follow. A consistent architecture and design framework will establish unknowns upfront, reduce the effort during the design phase of the individual migrations while also ensuring adherence to security, compliance, and other governance policies. It is worthwhile to mention that the CCOE team conducts the necessary Proof of Concepts at this stage to validate the critical architectural decisions.

*3) Build and Deploy Landing zone*

Every corporate application or service migrating to the cloud must adhere to the governance policies. A massive amount of time and effort will be wasted if every application team migrating to the cloud has to figure out how to stay compliant with the governance policies. Landing Zones prevent this wastage and accelerate cloud migration as they provide a baseline environment that is compliant with the governance policies. This covers key areas such as Identity and access management, network topology and connectivity, operations baseline, deployment methods, disaster recovery and business continuity. It is important to note that a Landing Zone is not a "done once and done" activity. As the cloud migration progresses, it needs to be managed, maintained, and tuned.

*4) Build Internal Tools to Support Cloud Migration*

Depending on the cloud model that an enterprise decides, the internal tools needed to support the migration and operations may differ. At a minimum, enterprises should build the necessary CI/CD (Continuous Integration / Continuous Deployment) pipelines to enable the build and deployment of software to the cloud. For hybrid or private cloud models, there could be a need to enhance the data analytics and monitoring tools.

*d) Change Management Workstream*

Focusing just on the technology shift will not lead to a successful cloud transformation. It is equally important, if not greater, to ensure that the people within the different areas of the organization are also given the necessary tools and means to embrace and accept changes that would be brought in by the cloud. It is important to note in this context that the emphasis should just not be on providing training on the new tools and technologies. Rather a comprehensive plan that also includes bringing a change in the culture - breaking silos within the organization, providing assurance to the staff that migrating to the cloud will not disrupt their jobs and that there are concrete career paths in the "new way of working" is extremely important. Regular open and transparent conversations by executives and senior leadership with the staff are essential to alleviate staff concerns and aid in them accepting and embracing change.

*D. Migrate*

The Migration workstream of the program plays a major role in this phase. The focus of this phase is to get applications migrated to the cloud. This phase starts with addressing 3 important questions: What is the migration strategy for each of the applications? What is the order in which applications should be moved to the cloud? What is the cut-over strategy for each of the applications? Answers to these questions establish the effort and schedule of the migrations.

An application migration strategy is an approach with which the services and their data are moved to the cloud. Generally, there are 7 strategies: refactor, replatform, repurchase, rehost, relocate, retain, and retire [12]. Generally, organizations that have many applications to migrate would want to sequence them so that lessons learnt from migrating the less critical applications can be used while migrating the more critical applications.

Once the key elements impacting migration efforts and schedule are clear, the next step is to establish plans for executing migrations. In cases where application migration is a simple lift-and-shift, a migration factory model can be established wherein a central team is established to migrate, validate and activate services in the cloud. However, in larger organizations with legacy applications, it is more often that such a simple lift-and-shift would not suffice, and appropriate migration strategies should be planned and executed.

*E. Operate and Optimize*

After an application migrates to the cloud, the focus should be to ensure that it is operating correctly and meeting the expectations of the customer as well as providing the expected business benefits. It is in this context that several actions are taken in the Foundation Phase of the program to help the organization operate the service in the cloud successfully. The training given to the operations personnel in the Change Management workstream ensures the readiness of the teams to address any customer complaints within the stipulated SLAs. Cloud Operations Management controls and tools provide teams with the ability to monitor the performance of the service, SLA compliance measurements. Cloud Financial Management controls and tools enable the teams to measure the costs being incurred for running the service in the cloud and that it is within the allocated budgets.

Once the service is operating stably in the cloud, it is time to think of optimizing it to leverage the benefits that are offered by cloud platforms. It is in this phase the teams start looking at how to leverage the elasticity and scalability benefits of the cloud, how to re-architect applications and processes to be able to use advanced features and services offered by the Cloud Service Provider.

## III. CONCLUSION

The proposed framework ensures that the critical success factors of executive buy-in and engagement, strong cloud strategy and cloud governance, effective change management and control of costs are weaved in. By using this framework, organizations can ensure their cloud migration initiatives succeed and allow them to leverage the full benefits that the cloud has to offer.